\documentclass[useAMS,usenatbib]{mn2e}
\usepackage[dvips]{graphicx,color}
\usepackage{txfonts}
\usepackage{amssymb}
\usepackage{indentfirst}
\usepackage{epsfig}

\newcommand{\MgXII}{\mbox{{Mg\,{\sevensize XII}}}}

\newcommand{\FeSi}{\ensuremath{\mbox{Fe~K}\alpha/\mbox{Si~K}\alpha~}}

\newcommand{\Lya}{\ensuremath{\hbox{Ly}\alpha~}}
\newcommand{\Lyb}{\ensuremath{\hbox{Ly}\beta~}}
\newcommand{\Ka}{\ensuremath{\hbox{K}\alpha~}}

\newcommand{\A}{\AA~}

\def\cha{{\it Chandra }}

\def\nus{{\it NuSTAR }}

\begin{document}
\title[The clumpy torus around type II AGN]{The clumpy torus around type II AGN as 
revealed by X-ray fluorescent lines} 

\author[J. Liu et al.]{Jiren Liu$^{1,}$\thanks{E-mail: jirenliu@nao.cas.cn},
	Yuan Liu$^{2}$, Xiaobo Li$^{2}$, Weiwei Xu$^{1}$, Lijun Gou$^{1}$, and Cheng Cheng$^{1}$\\
	 $^{1}$National Astronomical Observatories, 20A Datun Road, Beijing 100012, China\\
	 $^{2}$Institute of High Energy Physics, Chinese Academy of Science, P.O. Box 918-3,
	 Beijing 100049, China\\
}

\date{}

\maketitle

\begin{abstract}

The reflection spectrum of the torus around AGN is characterized by
X-ray fluorescent lines, which are most prominent for type II AGN.
A clumpy torus allows photons reflected from the back-side of the torus
to leak through the front free-of-obscuration regions. Therefore, the observed 
X-ray fluorescent lines are sensitive to the clumpiness of the torus.
We analyse a sample of type II AGN observed
with \cha HETGS, and measure the fluxes for the Si \Ka and Fe \Ka lines.
The measured \FeSi ratios, spanning a range between $5-60$, 
are far smaller than the ratios predicted from simulations of 
smooth tori, indicating that the tori of the studied sources have  
clumpy distributions rather than smooth ones.
Compared with simulation results of clumpy tori with a half-opening angle of 60$^{\circ}$,
the Circinus galaxy has a \FeSi ratio of $\sim60$, which is close to 
the simulation results for $N=5$, where $N$ is the average number of clumps along the line of sight.
The \FeSi ratios of the other sources are all below the simulation results for $N=2$.
Overall, it shows that the non-Fe fluorescent lines in the soft X-ray band are a potentially 
powerful probe of the clumpiness of the torus around AGN.

\end{abstract}


\section{Introduction}

The obscuring torus around active galactic nucleus (AGN) is the key ingredient
to unifying different types of AGN \citep[e.g.][]{Ant93}.
The distribution of matter in the torus was considered to be clumpy in
early theoretical studies
\citep{KB88}. This implies that the classification of AGN is 
a probability of direct view of AGN \citep{Eli08}.
The dust in the torus absorbs the radiation from the central AGN and re-radiates it in the infrared.
While the simplified smooth model of torus was studied from early on, many detailed infrared 
observations point to clumpy distributions \citep[see][for recent reviews]{Hon13,Net15}.

The clumpy nature of the torus is also supported by X-ray absorption variations 
on timescales from days to years. Different from dust emission in the 
infrared, obscuration of X-ray photons can be produced by dust-free gas within
the sublimation radius. An extreme example is NGC 1365, which shows an occultation
of the central AGN within 2 days, indicating that the obscuring clouds
are located at similar distances to those of the broad line region \citep[BLR,][]{Ris07} .
Statistical study of eclipse events in the {\it RXTE} (Rossi X-ray Timing Explorer) 
archive data by \citet{Mar14} 
showed that most of the clouds are located in the outer portion of the BLR.
Recent hard X-ray monitoring of NGC 1068 by \nus (Nuclear Spectroscopic Telescope Array) 
reveals flux variations above 20 keV,
which imply a transition of the column density from $10^{25}$\,cm$^{-2}$ to
$6.7\times10^{24}$\,cm$^{-2}$ \citep{Mar16}.
There seems to be a continuous distribution of clouds from the BLR to the parsec-scale torus
\citep[e.g.][]{Bia12,Net15}.

X-ray spectra of type II AGN with column densities larger than 
$10^{23}$\,cm$^{-2}$ are particularly suitable for the study of the torus, 
since their intrinsic continua are heavily obscured in the soft X-ray band.
This is especially true for Compton-thick AGN (with column densities 
$>1.2\times10^{24}$\,cm$^{-2}$), 
the intrinsic continua of which are heavily suppressed below 10 keV, and the observed spectra 
are dominated by reflection emission from the torus \citep[e.g.][]{Com04}. 
A characteristic feature of the reflection spectrum is the prominent
Fe K$\alpha$ emission line at 6.4 keV, which generally has an equivalent width (EW) $\sim1$ keV 
for Compton-thick AGN \citep[e.g.][]{Lev02}.

\begin{table*}
	\caption{List of the selected type II AGN observed with \cha HETGS}
\begin{tabular}{lccccccccc}
\hline
\hline
Name & I(Si K$\alpha$) & I(\MgXII\ Ly$\alpha$) & I$_c$(Si K$\alpha$)& I(Fe K$\alpha$) & I(Fe)/I$_c$(Si) & EW(Fe
K$\alpha$)&$N_
\rmn{H}^T$ & Reference of
$N_
\rmn{H}^T$  \\
\hline
Circinus & $3.7\pm0.7$ &$5.0\pm0.8$& $5.4\pm1.0$& $315.6\pm8.5$ &$58\pm11$&1.6 &8  &\citet{Are14} \\
Mrk 3    & $3.2\pm1.0$ &$3.2\pm0.9$& $3.4\pm1.1$& $50.2\pm4.9$  & $15\pm5$&0.46&0.9&\citet{Yaq15} \\
NGC 1068 & $6.4\pm1.5$ &$18.4\pm3.0$ & $6.4\pm1.5$& $43.6\pm5.6$  & $7\pm2$ &0.68&10 &\citet{Bau15} \\
NGC 3393 & $0.9\pm0.3$ &$0.8\pm0.3$& $0.7\pm0.3$& $3.7\pm1.5$  & $5\pm3$ &0.56&2.2 &\citet{Kos15} \\
NGC 7582 & $1.3\pm0.6$ &$1.6\pm1.0$& $1.0\pm0.6$& $19.2\pm4.3$ & $19\pm12$&0.39&3  &\citet{Riv15} \\
NGC 2992 & $2.5\pm0.9$ &$2.4\pm1.1$& $2.0\pm0.9$& $18.8\pm4.9$  & $9\pm5$ &0.43&2  & Xu et al. in prep\\
NGC 4388 & $2.1\pm0.5$ &$0.4\pm0.3$& $2.0\pm0.4$& $67.7\pm8.7$  & $34\pm8$&0.17&0.4&\citet{Shi08}\\
NGC 4507 & $1.7\pm0.6$ &$1.9\pm0.6$& $1.3\pm0.6$& $55.5\pm11.4$ & $43\pm22$&0.11&0.7&\citet{Mar13}\\
3C 445   & $0.7\pm0.2$ &$0.6\pm0.3$& $0.6\pm0.2$& $16.1\pm4.1$  & $27\pm11$&0.14&0.3&\citet{Bra11}\\
\hline
\end{tabular}
\begin{description}
\begin{footnotesize}
\item
  Note: 
  I is the line intensity in units of $10^{-6}$\,photons cm$^{-2}$ s$^{-1}$ measured from the 
  \cha HEG data.
  I$_c$(Si K$\alpha$) is the value corrected for Galactic absorption (Circinus and Mrk 3) and 
  the contribution of the \MgXII\ \Lyb line for other sources. 
  EW(Fe K$\alpha$) is in units of keV. $N_\rmn{H}^T$ (in units of 10$^{24}$\,cm$^{-2}$) 
  is the quoted column density of torus in recent literature.
  For NGC 2992, the Fe \Ka line of the \cha HEG spectrum has an EW $\sim0.4$\,keV, 
  which can be fitted with the scattered spectrum of the MYTorus model with a column density 
  $\sim2\times10^{24}$\,cm$^{-2}$ (Xu et al. in prep).
\end{footnotesize}
\end{description}
\end{table*}

A general feature of a clumpy torus is that soft X-ray photons, 
reflected from the back-side of the torus, are not significantly reduced due to 
their passage through the front free-of-obscuration regions between clumps \citep{NG94,Yaq12,Liu14}.
As shown by recent simulations, a clumpy torus provides significantly more 
soft X-ray photons than a smooth one when viewed edge-on, and the softer the photons, the larger
the differences \citep{Liu14}.
The reflection spectrum can therefore be used to 
probe the matter distribution of the torus. 
For example, \citet{Bau15} found that a multiple-component reflector is needed 
to fit the \nus spectra of NGC 1068. 
\citet{Bal14} found that the \nus spectra of 
NGC 424, NGC 1320, and IC 2560 are better fitted with a "face-on" component of 
MYTorus model \citep{MY09}, which was suggested to mimic the leakage of soft X-ray 
photons from a clumpy torus \citep{Yaq12}. 

As the soft X-ray spectra of AGN are generally contaminated by other 
components, such as emission from photo-ionized gas, 
it is hard to measure the pure reflected continuum 
in the soft X-ray band. Instead, the non-Fe fluorescent lines 
(e.g. Si \Ka line at 1.74 keV) emitted by the torus can be measured unambiguously,
since they are unique line features that can not be 
confused with other components. Because the fraction of leaked photons
depends on energy, the flux ratios of the Fe K$\alpha$ line to non-Fe \Ka lines would be a 
good probe of the matter distribution of the torus.
The non-Fe fluorescent lines are generally weak, due to their small yields 
and relatively low abundances \citep[e.g.][]{Rey94,Mat97}.
Nevertheless, they are still observable, especially for Compton-thick AGN when the intrinsic
continua are heavily suppressed \citep[e.g.][]{Xu16}. 

In this Letter we compile a sample of type II AGN, observed with \cha 
High Energy Transmission Grating Spectrometer (HETGS), 
to study their non-Fe fluorescent lines, focusing on the Si \Ka line at 1.74 keV, which is 
generally the strongest of the non-Fe fluorescent lines. 
We compare the measured flux ratios of \FeSi
with simulation results of clump tori by \citet{Liu14}.
The Fe K$\alpha$/Ni \Ka ratio has been investigated previously \citep[e.g.][]{Mol03,Mat04,YM11}.
Note that the errors quoted are for the 90\% confidence level.

\section{Observational data }

The \cha HETGS provides 
high resolution X-ray spectra for point-like sources \citep[][]{Can05}.
It consists of a High Energy Grating (HEG) with a spectral resolution of 
0.012 \A (full width half maximum, FWHM) and a Medium Energy Grating (MEG) with a 
resolution of 0.023 \AA. 
The effective area of MEG is about 3 times larger than that of HEG around 1.74 keV, and
drops quickly above 6 keV.
We searched the \cha HETGS archive for type II AGN having sufficient
signal-to-noise ratio (S/N) to allow a reliable measurement of the Si \Ka line.  
We found a total of eight AGN, which are listed in Table 1.
The Si \Ka line is also detected in Cen A, but 
its dust lane complicates the estimation of the real flux of the Si \Ka line and
is therefore not included. Although classified as Seyfert 1.5, 3C 445 has an X-ray spectrum 
similar to type II AGN \citep[e.g.][]{Sam07,Gra07} and has been included in the list.
The first five sources listed in Table 1 are generally regarded as Compton-thick AGN, while the 
others are less
obscured. Many of the selected sources have already been studied in the literature for their Fe \Ka line 
\citep[e.g.][]{Shu11}.

We use the data downloaded from \cha Transmission Grating Data Archive and Catalog \citep[TGCat,][]{TG}.
All the spectra are extracted from a region with a 2 arcsec half-width in the cross-dispersion direction,
and the $\pm1$ order data are combined together.
The foreground Galactic absorption column density is $6.2\times10^{21}$\,cm$^{-2}$ and
$1\times10^{21}$\,cm$^{-2}$ for Circinus and Mrk 3, respectively, and
is negligible ($<0.7\times10^{21}$\,cm$^{-2}$) for other galaxies \citep{Nh05}.

\begin{figure*}
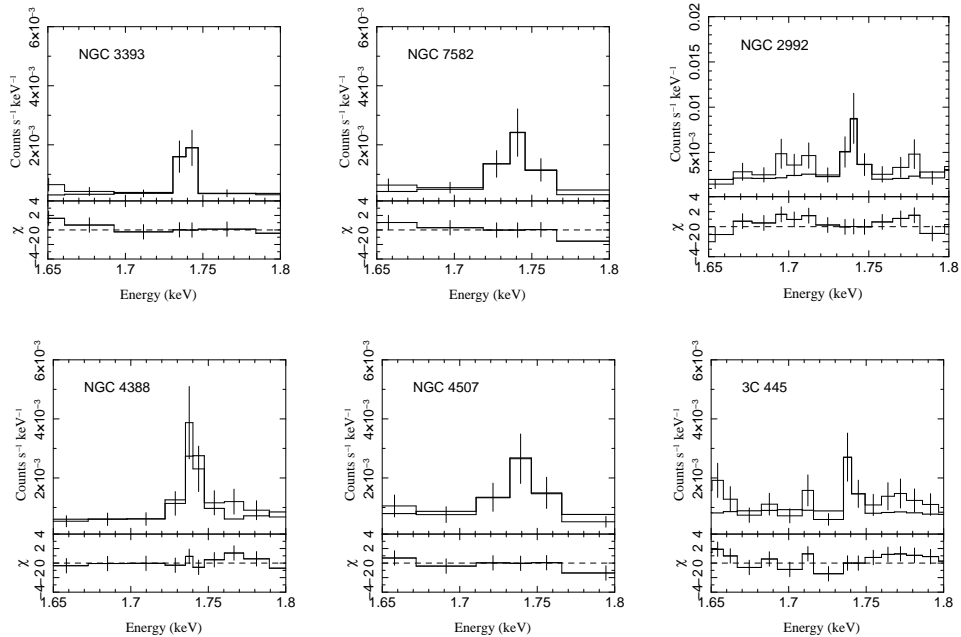

\hspace*{1.5cm}
\includegraphics[height=1.73in]{3393SiHH.ps}
\includegraphics[height=1.73in]{7582SiHH.ps}
\includegraphics[height=1.73in]{2992SiHH.ps}
\hspace*{1.5cm}
\includegraphics[height=1.73in]{4388SiHH.ps}
\includegraphics[height=1.73in]{4507SiHH.ps}
\includegraphics[height=1.73in]{445SiHH.ps}
\caption{
\cha HEG spectra of the Si \Ka line of the selected samples, except for Circinus, Mrk 3, and NGC 1068,
whose spectra were presented in \citet{Liu16}. 
The corresponding redshifts have been corrected. The data were binned to a minimum S/N ratio of 3.
The fitted models of one Gaussian line plus a power-law continuum are plotted as the thick 
solid lines.
}
\end{figure*}

\section{Results}

The Si \Ka line (at 1.74 keV) is adjacent to the \MgXII\ \Lyb line at
1.745 keV. For Circinus, Mrk 3, and NGC 1068, their \cha HEG data are deep enough
to resolve the peaks of these two lines, thus allowing the 
two lines to be measured simultaneously. In particular, the line width of the 
Si \Ka line can be measured with a spectral resolution 4 times 
better than that at 6.4 keV where the Fe \Ka line resides. 
Measurements of the Si \Ka line of
Circinus, Mrk 3, and NGC 1068 were presented in \citet{Liu16}. 
Si \Ka line widths are $3-5$ times smaller than those measured for
the Fe \Ka line, indicating that the line-emitting regions are outside the dust sublimation radius.
We refer the reader to \citet{Liu16} for more details and
for convenience, we have listed the relevant results in Table 1.

For all the other sources, the \cha HEG data are not deep enough to allow a deblending of
the Si \Ka line and the \MgXII\ \Lyb line.
The \cha HEG spectra of these sources are shown in Fig. 1.
It can be seen that the Si \Ka line and the \MgXII\ \Lyb line are blended with each other.
We fit the data with a model comprising one Gaussian line plus a power-law continuum. 
To improve the S/N ratio, the HEG and MEG data are fitted simultaneously.
The fitting region is between 1.4 and 2.3 keV and neglecting the other main emission features.
The fitted results are plotted in Fig. 1 and listed in Table 1.
We see that the one line model is sufficient to fit the data. 
If we replace the power-law continuum with a thermal continuum, the fitted intensities of 
Si \Ka line are generally consistent with those of the power-law model, 
with differences smaller than 2\%.

The contribution of the \MgXII\ \Lyb line can be estimated based on the fluxes of 
the \MgXII\ \Lya line at 1.472 keV, which are also listed in Table 1.
For Circinus, Mrk 3, and NGC 1068, the fluxes of the \MgXII\ \Lyb line can be measured
\citep{Liu16} and
the \MgXII\ Ly$\beta$/\MgXII\ \Lya ratios (corrected for Galactic absorption) 
are $0.38\pm0.11$, $0.69\pm0.33$, and $0.19\pm0.07$, respectively.
For photo-ionized plasmas, the emissivity
ratio between \MgXII\ Ly$\beta$ and \MgXII\ \Lya is around $0.1-0.25$ \citep[e.g.][]{Kin03}.
Considering the uncertainties in the measurement, the \MgXII\ Ly$\beta$/\MgXII\ \Lya ratios
of Circinus and NGC 1068 are consistent with photo-ionization models. For Mrk 3, the ratio
is larger than the predictions of photo-ionization models.
It is likely that the \MgXII\ Ly$\beta$ line is over-estimated due to blending 
with the Si \Ka line \citep{Liu16}.
Assuming a ratio of \MgXII\ Ly$\beta$/\MgXII\ \Lya of 0.2,
the contribution of the \MgXII\ \Lyb line to the measured fluxes of Si \Ka line is about 
20\% for all sources.

The fluxes of the Fe \Ka line are measured using only the \cha HEG data. We also adopt the model of
one Gaussian line plus a power-law continuum. The fitting region is between 5 and 6.8 keV, and
the measured fluxes are listed in Table 1. For reference, the EWs of the samples measured 
from the fitted models are also included. The measured EWs are generally larger than
0.3 keV, indicating the reflection-dominated nature of the X-ray spectra of the sample.
From the measured fluxes of the Si \Ka and Fe \Ka line, we 
can calculate the \FeSi ratio after correcting for the contribution of the 
\MgXII\ \Lyb line. The results are listed in Table 1.
We see that they span a range between $5-60$. 
The Circinus galaxy has the largest \FeSi ratio around 60, 
while the ratios of NGC 1068, NGC 3393, and NGC 2992 are as small as $\sim7$.

\citet{Liu14} simulated the reflection spectrum of a torus with clumpy distributions. 
The boundaries of the torus
are defined by an inner radius of 0.1 pc, an outer radius of 2 pc, and a half-opening angle
of 60$^\circ$. This corresponds to a covering factor of 0.5 for a smooth torus.
The clumpiness of the torus is described by the volume filling factor,
the average number of clouds along the line of sight ($N$), and the average column density ($N_\rmn{H}$).
The smaller the $N$, the more clumpy the torus.
\citet{Liu14} found that the filling factor only slightly affects the spectra, while $N$ can change the emergent
spectra significantly.
A crude grid of $N_\rmn{H}$ (10$^{23}$, 10$^{24}$ and 10$^{25}$\,cm$^{-2}$) and $N$ (2, 5, 10) 
was simulated for a fixed photon index of $\Gamma=1.8$. 
The simulated \FeSi ratios for $N_\rmn{H}=10^{24}$\,cm$^{-2}$ and 10$^{25}$\,cm$^{-2}$ 
with a filling factor of 0.01 are plotted in Fig. 2.
For comparison, the results of the smooth cases are also plotted.

\begin{figure}
\includegraphics[height=2.0in]{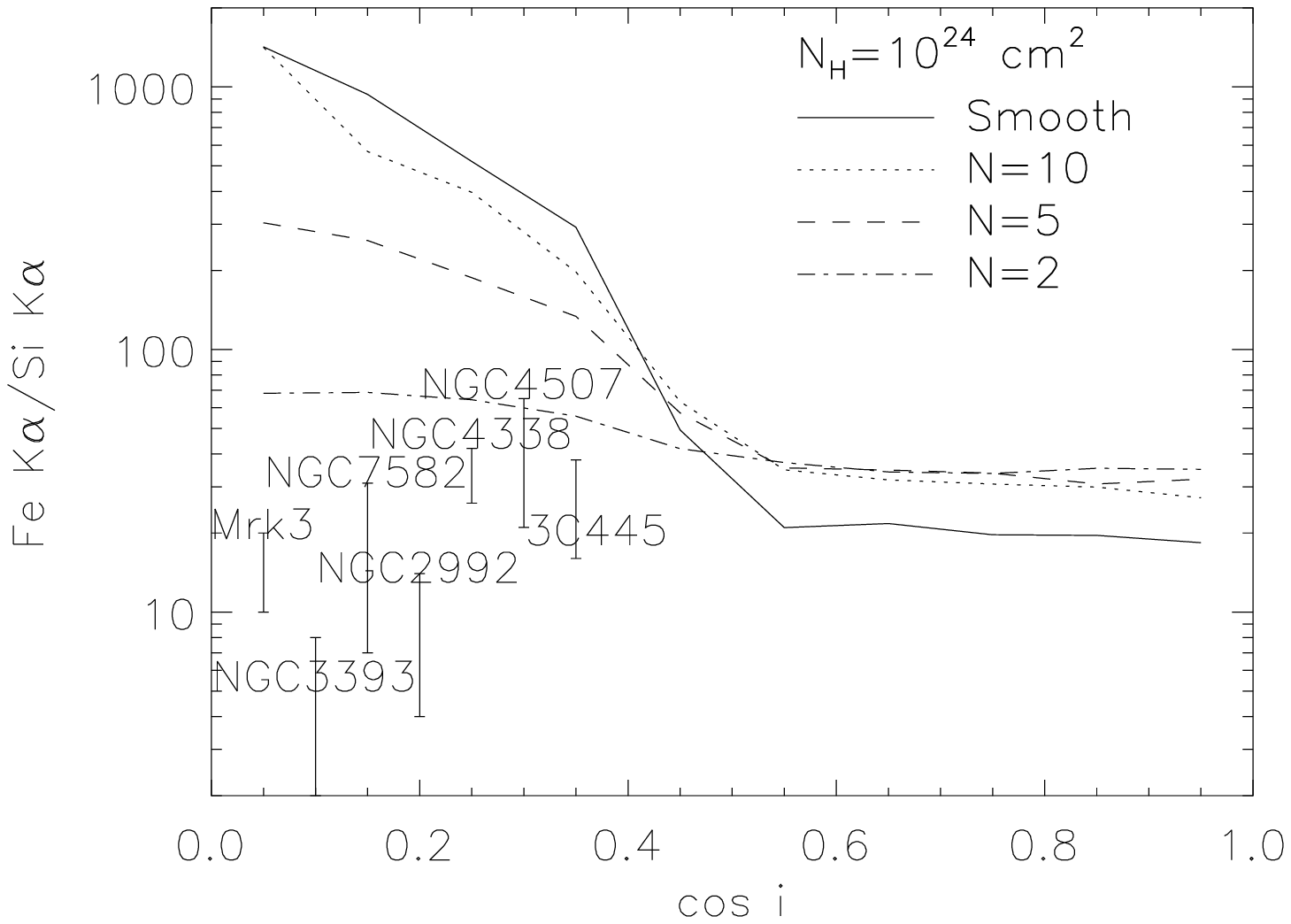}
\includegraphics[height=2.0in]{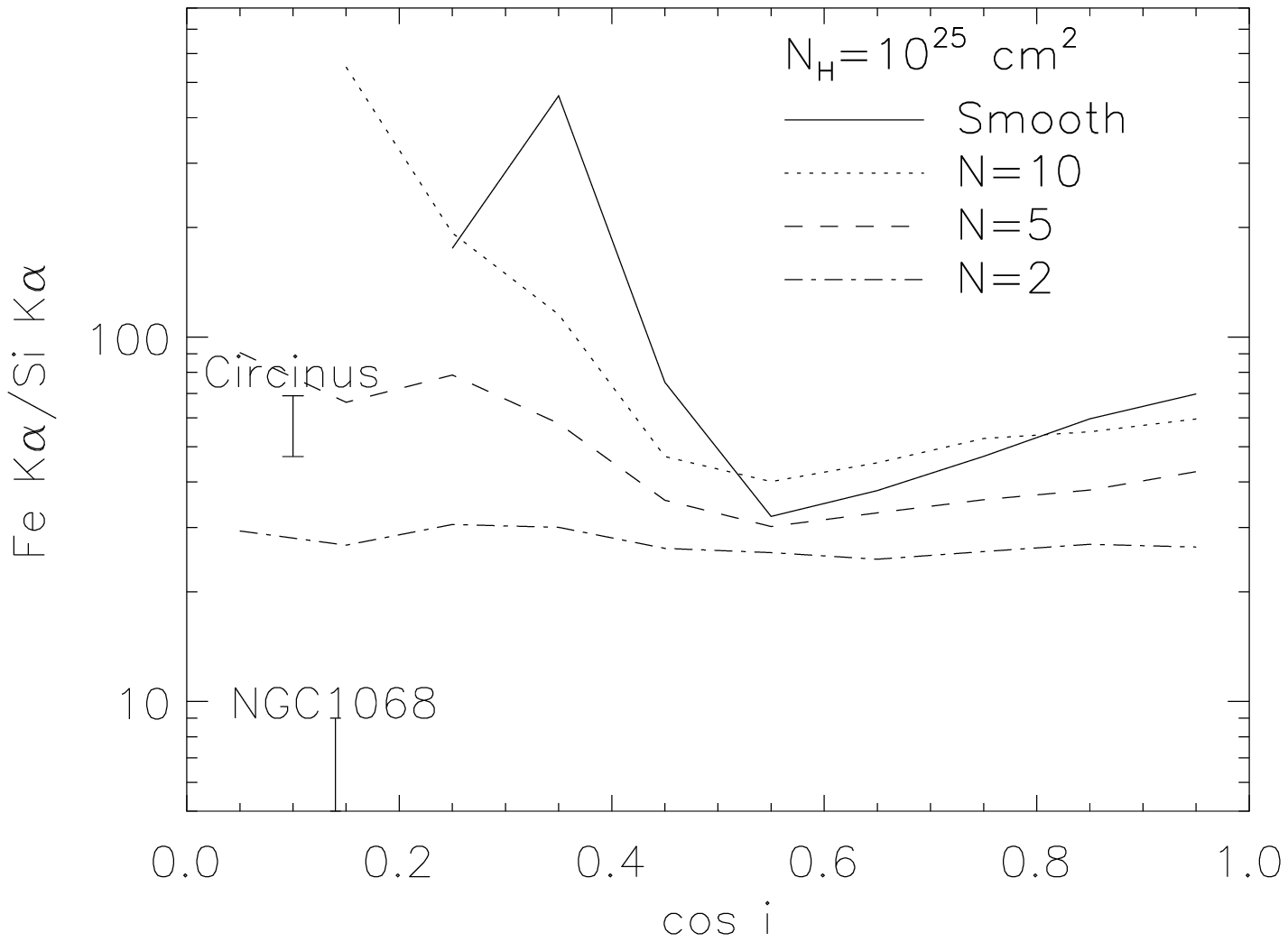}
\caption{\FeSi ratio against cosine of the inclination angle from simulations of 
	clumpy tori with $N_\rmn{H}=10^{24}$\,cm$^{-2}$ and 10$^{25}$\,cm$^{-2}$ by \citet{Liu14}. 
	The inclination angle 0$^\circ$ corresponds to face-on viewing, while 90$^\circ$ to edge-on.
	$N$ is the average number of clouds along the line of sight, which is smaller for
	a more clumpy torus.
	The measured ratios are over-plotted according to their column densities, with 
	an inclination angle arbitrarily assigned.
	Note that the measured ratios are all far below the predictions of the smooth torus,
	indicating that the tori of the samples have a clumpy distribution.
}
\end{figure}

First of all, we see that for the torus of $N_\rmn{H}=10^{24}$\,cm$^{-2}$ with a smooth distribution, 
the \FeSi ratio becomes smaller when the viewing angle changes from edge-on to face-on.
This is due to less absorption of Si \Ka photons in the face-on direction. For the clumpy torus,
the \FeSi ratio follows the same trend, but with a smaller value when viewed 
edge-on. The more clumpy the torus, the smaller the ratio. 
This is due to
leakage of Si \Ka photons as discussed in the Introduction.
For the case of $N_\rmn{H}=10^{25}$\,cm$^{-2}$, 
the absorption becomes so severe that even no Si \Ka photons appeared on some sight-lines. 
Nevertheless, the differences between the clumpy and smooth
torus are similar to those for $N_\rmn{H}=10^{24}$\,cm$^{-2}$. 
The torus of $N_\rmn{H}=10^{25}$\,cm$^{-2}$ 
is optically thick 
for Fe \Ka photons, and the absorption of Fe \Ka photons leads to
a smaller \FeSi ratio when viewed edge-on compared with the $N_\rmn{H}=10^{24}$\,cm$^{-2}$ case. 
Note that, for $N=2$, the \FeSi ratios become almost isotropic, the effect of which
was discussed in detail in \citet{Liu14}.

The measured \FeSi line ratios are over-plotted in Fig. 2, either 
in the $N_\rmn{H}=10^{24}$\,cm$^{-2}$ or the $10^{25}$\,cm$^{-2}$ panel, according to
the column densities quoted in recent literature (shown in Table 1). 
The viewing angle is assigned with an arbitrary value larger than the half-opening
angle of the simulated geometry.
For the type II AGN studied here, to obscure the central AGN, the viewing angle 
must be larger than the half-opening angle of the torus. 
For Compton-thick sources, the reflection-dominated spectra indicate that
they are most likely to be viewed edge-on.
We see that all the measured \FeSi ratios
are far smaller than the predictions of the smooth torus with viewing angles larger than
the half-opening angle of the torus.
This indicates that the distribution of the tori of the sources is far from being smooth.
For the simulated geometry of the clumpy torus, only the \FeSi ratio of the Circinus galaxy is 
close to the $N=5$ simulation results. For the other sources, the measured ratios are all
below the simulation curve fir $N=2$.

\section{Conclusion and Discussion}

X-ray fluorescent lines are unique features of the reflection spectrum 
of the torus around AGN, and can be relatively easily detected 
for obscured type II AGN.
The spectral shape of the reflection spectrum depends on the fraction
of photons that can leak through the torus, and thus on the
clumpiness of the torus. 
Therefore, X-ray fluorescent lines at different energies
can be used to probe the clumpiness of the torus.
We have analysed a sample of type II AGN observed
with \cha HETGS and measured the fluxes for the Si \Ka and Fe \Ka lines.
The measured \FeSi ratios span a range between $5-60$.
These ratios are far smaller than the simulated values for smooth tori,
indicating that the tori of the samples have clumpy distributions, 
rather than smooth ones.

\begin{figure}
\includegraphics[height=2.0in]{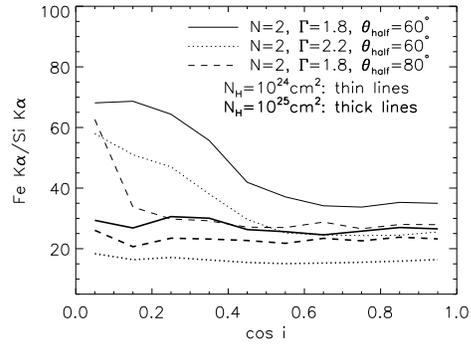}
\caption{The effects of the photon index $\Gamma$ and the half-opening angle of the torus
	on the predicted \FeSi ratio.
}
\end{figure}

Compared with the simulation results of clumpy tori,
the Circinus galaxy has a \FeSi ratio of $\sim60$, which is close to the values
of the $N=5$ simulation . The \FeSi ratios of the other sources are all
below the results of the $N=2$ simulation. This implies a very clump distribution for the torus.
Below we discuss several factors that can affect the simulation results.

A direct factor affecting the simulation results is the abundance pattern.
\citet{Liu14} adopted the solar abundances from \citet{AG89}, and no variable 
abundances are investigated. Recent studies have reduced the solar Fe abundance
from $4.68\times10^{-5}$ \citep[number relative to H,][]{AG89} to 
$\sim3\times10^{-5}$ \citep[e.g.][]{Asp09}.
This will lead to a similar decrease in the predicted \FeSi ratio if the 
current abundance pattern is assumed. Similarly, if the abundance ratio of Fe/Si is changed
several times, the \FeSi ratio will also change similarly.

Another uncertainty in the results is the photon index, $\Gamma$, which is fixed at 
1.8 in \citet{Liu14}. For optically-thick material, the emission of fluorescent lines
is proportional to the available number of photons that can ionize the 
corresponding K-shell electrons. The relative number of photons at the ionization edge 
of Si and Fe will be increased for a large $\Gamma$. 
To test the effect of $\Gamma$, we re-run the $N=2$ simulation 
using a $\Gamma$ of 2.2. The results for $N_\rmn{H}=10^{24}$\,cm$^{-2}$ 
and $10^{25}$\,cm$^{-2}$ are plotted
in Fig. 3 as dotted lines. As can be seen, for $N_\rmn{H}=10^{25}$\,cm$^{-2}$,
the \FeSi ratios are reduced by a 
factor of $\sim1.7$, just as expected for optically-thick material. 
For $N_\rmn{H}=10^{24}$\,cm$^{-2}$, 
the material is not optically-thick for Fe K-shell ionization yet, and the reduction in the 
ratio is less than the prediction for optically-thick material.

Another unexplored effect is the half-opening angle (or covering factor), which is fixed
to be 60$^\circ$. In principle, a smaller covering factor will allow more photons to
reach the observer, since it means less obscuration by the front-side torus gas. The effect on
soft photons would be more significant due to their larger absorption cross-sections.
Thus, smaller \FeSi ratios would be expected for smaller covering factors.
To test the effect of covering factor, we also re-run the $N=2$ simulation 
by adopting a half-opening angle of 80$^\circ$, the results of which are plotted in 
Fig. 3 as dashed lines. For $N_\rmn{H}=10^{24}$\,cm$^{-2}$, the \FeSi ratio can be decreased by a factor of 2 
for edge-on angles when using a half-opening angle of 80$^\circ$.
For $N_\rmn{H}=10^{25}$\,\,cm$^{-2}$, the effect of changing the half-opening angle is less significant, 
due to severe absorption and the isotropic nature of the emission.

We note that the simulated column densities (10$^{24}$ and 10$^{25}$\,cm$^{-2}$) are not matched
exactly with the observed ones of the torus that were determined from the smooth torus models, 
which are shown here to be invalid.
It seems that by including all these factors and with a clumpy torus of $N\sim2$ or smaller, the 
measured relatively low \FeSi ratios ($\sim7$) can be obtained, without resort to 
abnormally low Fe/Si abundances. 
Current simulations of clumpy tori are limited by the crude grid and 
fixed abundances, geometry, and photon spectrum.
It is clear that a fittable clumpy model, including the effects of $N_\rmn{H}$, $\Gamma$, 
variable abundances and covering factors, is needed to fully understand the X-ray 
spectra of obscured AGN and to constrain the clumpiness of the torus.
Such a model will be presented in a future work.

The simulation of \citet{Liu14} assumed that both the Fe \Ka and Si \Ka photons
are from the same torus, with the Si \Ka photons from a thinner layer due to the 
larger absorption section at lower energy. In principle, the Si \Ka photons
can arise from physically distinct clouds with $N_\rmn{H}\sim10^{22}$\,cm$^{-2}$,
for which the \FeSi ratio is reduced significantly. For sources with 
$N_\rmn{H}>10^{24}$\,cm$^{-2}$ and a covering factor $\sim0.5$, the contribution
of the Si \Ka photons from distinct clouds with $N_\rmn{H}\sim10^{22}$\,cm$^{-2}$
is unlikely to be significant, since the intensity of Si \Ka photons increases with
column density until saturated for large optical depth. For the extreme case
that clouds with $N_\rmn{H}\sim10^{22}$\,cm$^{-2}$ cover another half sky, the 
reduction of the \FeSi ratio can not be larger than 50\%. Nevertheless, the possibility
of the origin of Si \Ka photons from distinct clouds could be important for 
sources with $N_\rmn{H}\sim10^{23}$\,cm$^{-2}$.

Considering the contamination by the \MgXII\ \Lyb line to the Si \Ka line, the relatively 
isolated S \Ka line at 2.3 keV would be a better probe. However, due to the quick decline
of the effective area of HEG around 2 keV, we found the S \Ka line is only detectable in 
Circinus, Mrk 3, and NGC 1068, and the flux is similar to that of the Si \Ka line.
Nevertheless, the effective area around S, Ar, and Ca \Ka lines
would be improved for the soft X-ray calorimeter on-board Astro-H, which would
also enable a much larger sample of obscured AGN. 
The results show that non-Fe fluorescent lines are a potentially powerful probe of 
the clumpiness of the torus around AGN.

\section*{Acknowledgements}
We thank our referee Franz Bauer for pointing out the possible distinct origin
of Si \Ka photons and other valuable suggestions. We also thank Richard Long for a 
careful reading of the manuscript.
JL is supported by NSFC grant 11203032 and YL supported by NSFC grant 11573027.
This research is based on data obtained from the \cha Data Archive.

\bibliographystyle{mn2e}

\end{document}